\documentclass[floats,floatfix,showpacs,amssymb,prd,twocolumn,superscriptaddress,nofootinbib]{revtex4-1}

\usepackage{amssymb,amsmath,verbatim,mathtools,needspace,enumitem,etoolbox,graphicx}

\usepackage[dvipsnames, usenames]{xcolor}

\definecolor{linkcolor}{rgb}{0.0,0.3,0.5}
\usepackage[unicode, colorlinks=true, linkcolor=linkcolor, citecolor=linkcolor, filecolor=linkcolor,urlcolor=linkcolor, pdfusetitle]{hyperref}
\usepackage[all]{hypcap}
\usepackage[T1]{fontenc}
\usepackage[utf8]{inputenc}

\usepackage{epstopdf}

\usepackage{bm}
\usepackage{dcolumn}
\usepackage{longtable}
\usepackage{enumerate}
\usepackage[normalem]{ulem}

\def\be{\begin{equation}}
\def\ee{\end{equation}}
\newcommand{\beq}{\begin{eqnarray}}
\newcommand{\eeq}{\end{eqnarray}}

\newcommand{\tn}{\textnormal}

\begin{document}
{\raggedleft CERN-TH-2019-070 \\}
{\raggedleft CP$^\textnormal{3}$-Origins-2019-20 DNRF90 \\}

\title{The Photon Spectrum of Asymmetric Dark Stars}
\author{Andrea Maselli}\email{andrea.maselli@roma1.infn.it}
\affiliation{Dipartimento di Fisica, Sapienza Universit\'a  di Roma \& Sezione INFN Roma1, P.A. Moro 5, 00185, Roma, Italy}

\author{Chris Kouvaris}
\email{kouvaris@cp3.sdu.dk}
	\affiliation{CP$^3$-Origins, Centre for Cosmology and Particle Physics Phenomenology University of Southern Denmark, Campusvej 55, 5230 Odense M, Denmark}
	\affiliation{Theoretical Physics Department, CERN, 1211 Geneva, Switzerland}

\author {Kostas D. Kokkotas}\email{kostas.kokkotas@uni-tuebingen.de}
\affiliation{Theoretical Astrophysics, IAAT, University of T\"uebingen, 
T\"uebingen 72076, Germany}

\date{\today}
\begin{abstract}
Asymmetric Dark Stars, i.e., compact objects formed from the collapse of asymmetric 
dark matter could potentially produce a detectable photon flux if dark matter particles 
self-interact via dark photons that  kinetically mix with ordinary photons. The morphology 
of the emitted spectrum is significantly different and therefore distinguishable from a 
typical black-body one. Given the above and the fact that asymmetric dark stars can 
have masses outside the range of neutron stars, the detection of such a spectrum can 
be considered as a smoking gun signature for the existence of these exotic stars.
\end{abstract}

\maketitle 

\section{Introduction}\label{Sec:intro}
Today there is strong evidence about the existence of dark matter (DM) from a variety of 
different sources and scales, such as rotational curves of individual galaxies, clusters of 
galaxies such as the bullet cluster~\cite{Clowe:2006eq}, and the Cosmic Microwave 
Background~\cite{Ade:2015xua}. The so-called Collisionless Cold Dark Matter (CCDM) 
paradigm is well consistent with the observed large scale structure of the Universe. However, 
this picture changes at small scales. CCDM suffers from several issues that have currently 
not been resolved within the CCDM scenario. One of them is the {\it core-cusp} problem of 
dwarf galaxies, which is related to the fact that dwarf galaxies are observed to have flat density 
profiles in their central regions \cite{Moore:1994yx,Flores:1994gz}, while cuspy profiles for 
collisionless DM are predicted by N-body simulations \cite{Navarro:1996gj}. The latter also 
predict dwarf galaxies with  masses  that are too large to have not produced stars inside. 
However such dwarf galaxies  have not been observed yet \cite{BoylanKolchin:2011de}, leading 
to the so called {\it too-big-to-fail} problem. Furthermore a diversity problem does exist: galaxies 
with the same maximum velocity dispersion can differ significantly in the inner core~\cite{Oman:2015xda,deNaray:2009xj}, 
in contrast with the results obtained from N-body CCDM simulations. The resolution of such 
problems can be attributed to different factors: e.g. the inclusion of the baryonic feedback could 
potentially resolve the core-cusp problem, while statistical deviations from mean values could 
explain the too-big-to-fail problem. However, an attractive alternative solution to the CCDM 
inconsistencies is the existence of DM self-interactions (see~\cite{Tulin:2017ara} and reference 
therein for a review). Self-interactions of the order of $1\text{cm}^2/\text{g}$ are sufficient, for 
example, to disperse DM particles and flatten the density profile in dwarf galaxies. 

Moreover, DM self-interactions could resolve another potential problem related to supermassive 
black holes at high redshifts. Present models of stellar evolution do not seem  able to explain 
how these objects can grow to their current mass within the age of the Universe, if they start as 
typical stellar remnants. However, collapsing DM can provide the seeds for such supermassive 
black holes~\cite{Pollack:2014rja}.

Given the above, if DM possesses sufficiently strong self-interactions and is of asymmetric 
nature, it could potentially form compact objects. Asymmetric DM is a viable alternative to the 
thermally produced DM scenario. An asymmetry in the population between DM particles and 
antiparticles in the early Universe can lead to the depletion of the antiparticles via annihilations, 
resulting  to the survival of the component in excess. Once annihilations deplete the antiparticle 
population, no further annihilations can take place simply because DM consists only of particles 
and not antiparticles. Therefore in such a case, collapsing DM with the appropriate self-interactions  
needed to evacuate efficiently the energy, will result in the formation of neutron star-like dark matter 
objects. The possibility of asymmetric DM forming ``dark stars" was first explored in the context of 
fermionic~\cite{Kouvaris:2015rea} and bosonic DM~\cite{Eby:2015hsq} where star profiles and basic 
properties were determined. In addition, the possibility of forming admixed DM-bayonic 
stars~\cite{Tolos:2015qra,Deliyergiyev:2019vti} or asymmetric DM cores inside neutron stars 
(NS)~\cite{Kouvaris:2018wnh,Gresham:2018rqo} also exists. In standard cosmological scenarios, 
the formation of asymmetric dark stars requires an efficient energy evacuation mechanism. Such a 
mechanism was described and studied in~\cite{Chang:2018bgx}, in which asymmetric DM is assumed 
to feature self-interactions mediated by a dark photon.Collapsing DM can evacuate energy and shrink 
further via dark Bremsstrahlung. In this paper was  also shown that the collapsing  DM cloud can 
fragment and eventually form dark stars of different masses (depending on the DM and mediator 
masses and the dark photon coupling).

Asymmetric dark stars could in principle be detected via gravitational waves produced from mergers 
of such objects. The signal can be distinguished from similar mass black hole 
binaries~\cite{Cardoso:2017cfl,Maselli:2017cmm,Cardoso:2017cqb,Barack:2018yly,Cardoso:2019rvt,Pani:2019cyc} 
or NS ones~\cite{Maselli:2017vfi}, exploiting the effect of tidal interactions during the orbital evolution.

Although it seems that it will be difficult to detect dark stars without gravitational wave observations, in this 
paper here we investigate another possibility given by the emission of a photon flux. Our model is simply 
described by the following Lagrangian:
\begin{align}
\mathcal{L} =\bar{X}  &{\gamma}^\mu D_\mu {X}
-
m_\tn{X}
\bar{X} {X}
-
 \frac{1}{4}
F'_{\mu \nu}F'^{\mu \nu}\nonumber\\
&+
\frac{1}{2}m_{A'}^2 A'_\mu A'^\mu +\frac{1}{2}m_{D}^2 A_\mu A^\mu +\frac{\kappa}{2} F'_{\mu \nu}F^{\mu \nu}+...\ ,
\label{eq:lagrangian}
 \end{align}
where $X$ is the asymmetric DM particle, which couples to a dark photon $A'_{\mu}$ of mass $m_{A'}$, 
and possesses a kinetic mixing with the ordinary photon $\kappa$. We assume that the dark photon acquires 
its mass via an unspecified Higgs mechanism. The details of this mechanism are not important for the 
purposes of this work. $D_{\mu}$ is the usual covariant derivative $D_{\mu}=\partial_{\mu}-igA'_{\mu}$ ($g$ being the coupling to the dark photon). We have omitted the Standard Model part of the Lagrangian.
Dark/hidden photons mixing with ordinary photons have been studied thoroughly in literature especially in the context of their  effect on stars~\cite{An:2013yfc,Redondo:2013lna,Arias:2012az,Ayala:2019isl,Chu:2019rok,DeRocco:2019njg}. 
Eq.~(\ref{eq:lagrangian}) contains also a mass term for the ordinary photon. Although this is unacceptable in the vacuum,
we will discuss later on how such a mass term can arise in the medium.
In the absence of the aforementioned mass term in eq.~(\ref{eq:lagrangian}), by shifting appropriately the photon field $A_\mu$, one 
can eliminate the kinetic mixing term from the Lagrangian, inducing a suppressed coupling between dark 
photon and charged Standard Model (SM) particles, while leaving the dark sector intact, i.e., no coupling 
between $X$ and $A_{\mu}$. However we will argue later that in the context of a dark star, there are ways 
to generate such a photon mass term. Shifting the photon field will induce a mixed mass term $\kappa m_D^2 A'_{\mu}A^{\mu}$ which upon diagonalization of the
mass matrix gives
\begin{align}
\mathcal{L} =\bar{X}  {\gamma}^\mu D_\mu {X}&
-
m_\tn{X}
\bar{X} {X}
-
 \frac{1}{4}
F_{\Phi\mu \nu}F_{\Phi}^{\mu \nu}  -
 \frac{1}{4}
F_{\Phi'\mu \nu}F_{\Phi'}^{\mu \nu} , 
\label{eq:lagrangian2}
 \end{align}
where $\Phi$ and $\Phi'$ are the new diagonal fields which are linear combinations of the old $A$ and $A'$. The covariant derivative now reads $D_{\mu}=\partial_{\mu}-ig \Phi'_{\mu}-ig\kappa m_D^2/(m_{A'}^2-m_D^2)\Phi_{\mu}$. To leading order $A_{\mu}\simeq \Phi_{\mu}$ and therefore SM particles couple mainly to  $\Phi$. Therefore there is an effective coupling of the DM $X$ to ordinary photons $\sim g\kappa m_D^2/(m_{A'}^2-m_D^2)$.
 Under these assumptions, asymmetric dark stars can 
produce a spectrum of detectable photons via dark Bremsstrahlung. 

Although the overall luminosity of such objects is suppressed by a power of $ \kappa$, its magnitude 
could still be significant, and not dramatically smaller than the luminosity of a standard NS. This somewhat 
unexpected result can be easily understood. Photons can be produced almost anywhere in the bulk of a 
dark star and, as long as their mean free path is larger than the stellar radius, they can escape. On the 
contrary, photons produced by NS are exclusively emitted from their surface, as the mean free path is 
very small compared to the radius. This {\it volume vs surface} effect has also another dramatic impact 
on the observed flux. Photons produced at different radii inside a dark star are redshifted, as they escape, 
by a different amount due to the gravitational potential. As we show in the next sections, these features 
lead to a very peculiar photon spectrum, {\it that is  completely different than black-body radiation, and 
thus represent a unique smoking gun signal for the discovery of asymmetric dark stars.}

Throughout the paper we use natural units, in which $c=\hbar=k_\tn{B}=1$.

\section{Dark stars structure}\label{Sec:EoS}

In this work we consider fermionic asymmetric dark stars. We summarise here the basic features 
of the DM equation of state (EoS), referring the reader to \cite{Maselli:2017vfi} 
(and reference therein) for more details on the stellar models and  their macroscopic 
properties.

The fermionic DM particles  of mass $m_\tn{X}$ feel the repulsive dark photon interaction which due to 
the dark photon mass $m_{A'}$ behaves as a Yukawa potential of the form
\begin{equation}
	V = \frac{\alpha_\tn{X} }{r} e^{- m_{A'} r} ,\label{Yukawa}
\end{equation}
with  $\alpha_\tn{X}$ being the dark fine structure constant. The stellar pressure originates from Fermi 
exclusion principle and from the short-range interaction \eqref{Yukawa}. Pressure and energy density are 
computed in the mean field approximation \cite{Kouvaris:2015rea} as a function of the DM Fermi 
momentum $x=p_\tn{F}/m_\tn{X}$: 
\begin{subequations}
	\begin{align}
	\rho &= m_\tn{X}^4 \left[\frac{\xi (x)}{8\pi^2} + \frac{2}{9\pi^3}\alpha_\tn{X}\frac{ m_\tn{X}^2}{m_{A'}^2} x^6\right]\ ,\label{Eq: Fermion EoS1}\\
	p &= m_\tn{X}^4 \left[\frac{\chi (x)}{8\pi^2} + \frac{2}{9\pi^3}\alpha_\tn{X}\frac{ m_\tn{X}^2}{m_{A'}^2} x^6\right]\ .\label{Eq: Fermion EoS2}
\end{align}
\end{subequations}
The two functions $\xi$ and $\chi$ encapsulate the effect of the Fermi-repulsion 
\cite{1983bhwd.book.....S} and read:
\begin{align*}
	\xi(x) &= x\sqrt{1+x^2}(2x^2+1)- \ln\left(x+\sqrt{1+x^2} \right) ,\\
		\chi(x) &= x\sqrt{1+x^2}(2/3x^2-1)+ \ln\left(x+\sqrt{1+x^2} \right) .
\end{align*}
At low (high) densities eqns.~\eqref{Eq: Fermion EoS1}-\eqref{Eq: Fermion EoS2} 
reduce to a polytropic EoS $P = K \rho^\gamma$ with index 
$\gamma \simeq 5/3$ ($\gamma \simeq 1$).

\begin{figure*}[th]
\centering
\includegraphics[width=8cm]{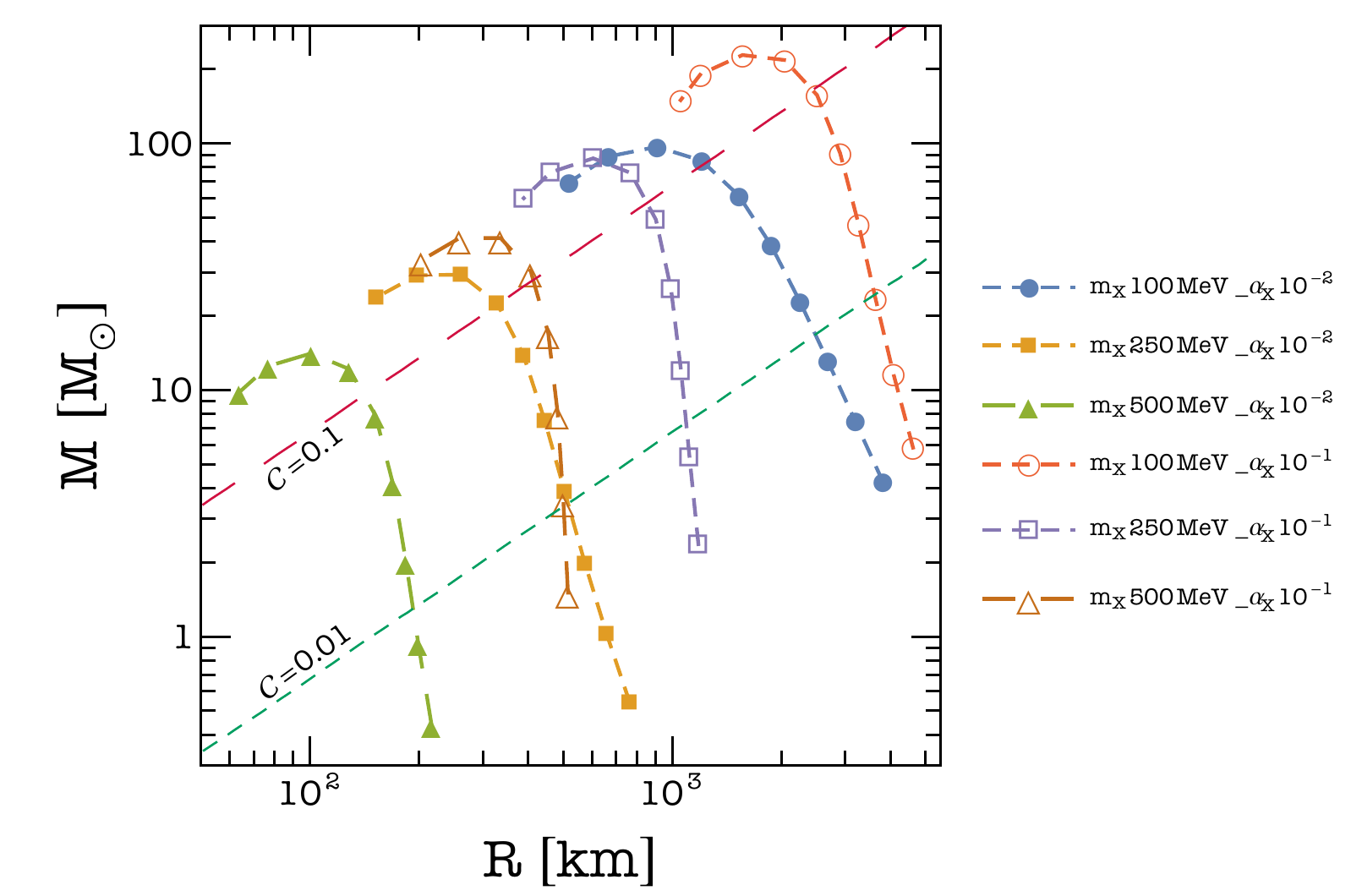}
\includegraphics[width=8cm]{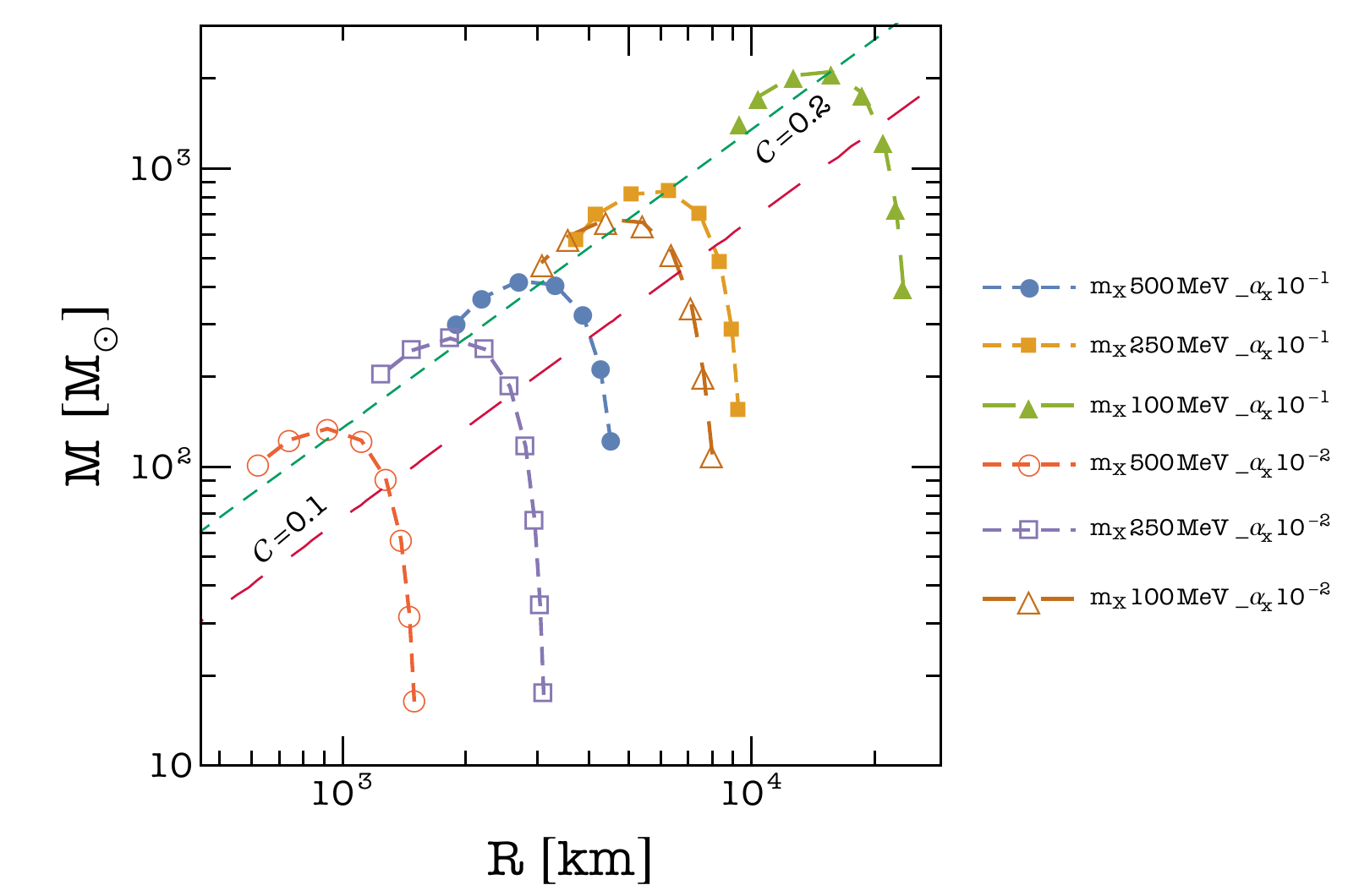}
\caption{Mass-radius profiles for the fermionic 
EoS. The configurations are identified by the dark 
photon mass $m_{A'}=10$MeV (left panel) and  
$m_{A'}=1$MeV (right panel), by the DM mass 
$m_\tn{X}=(100,250,500)$MeV, and by the constant 
$\alpha_\tn{X}=(10^{-1},10^{-2})$. Dashed 
straight lines identify stellar configurations 
with constant compactness ${\cal C}=M/R$.} 
\label{fig:MRprofiles} 
\end{figure*}

To build the star's structure, we solve Einstein's equations for a spherically 
symmetric, stationary and static spacetime, supplied by the fermionic EoS. 
We assume the following ansatz for the non-rotating star:
\begin{align}
ds^2&=g_{\mu\nu}dx^\mu d x^\nu\nonumber\\
&=-e^{2\nu(r)}dt^2+e^{2\lambda(r)}dr^2+r^2 d\Omega^2\ ,\label{ds2}
\end{align}
in the Schwarzschild coordinates $x^\mu=(t,r,\theta,\phi)$, where 
$e^{-2\lambda(r)}=1-2Gm(r)/r$. The latter leads to the relativistic stellar 
equations:
%
\begin{subequations}
\begin{align}
\frac{dp}{dr} &= -G\frac{(\rho+ p)\left[m(r)+ 4\pi r^3 p \right]}{r[r-2Gm(r)]}\ ,\\
\frac{dm}{dr} &= 4\pi r^2\rho \ ,\\
\frac{d\nu}{dr} &= -\frac{1}{p+\rho}\frac{dp}{dr} \ , 
\end{align}
\end{subequations}
which we solve with appropriate boundary conditions at 
the center of the star, up to the surface, where 
$r=R$, $m(r)=M$ and $p(R)\rightarrow0$. The mass-radius 
profiles of all the configurations considered 
in this paper, are shown in Fig.~\eqref{fig:MRprofiles}, 
for different values of the DM mass, 
$m_{A'}=(10\tn{MeV},1\tn{MeV})$ and  
$\alpha_\tn{X}=(10^{-1},10^{-2})$.

\section{Bremsstrahlung emission}\label{Sec:brems}

\begin{figure}[h]
\centering
\includegraphics[width=6cm]{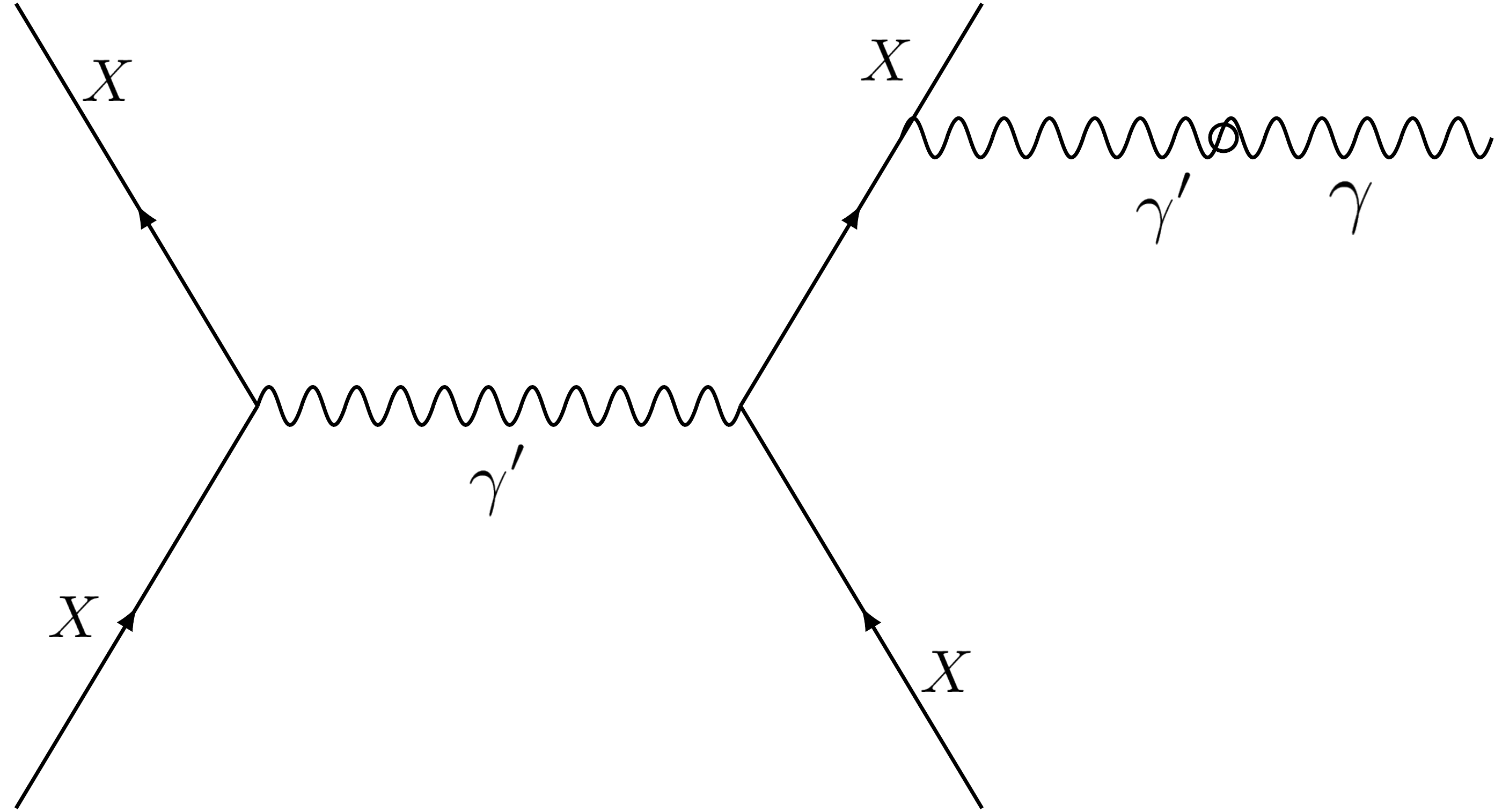}
\caption{Feynmann diagram of the Bremsstrahlung process 
we consider in this paper. Note that the photon can be 
attached to any DM leg. In the diagonal basis the DM exchange a $\Phi'$ and emit a $\Phi$.} 
\label{fig:diagram}
\end{figure}

We consider the emission of Bremsstrahlung photons with  
momentum $k^\mu=(\omega,\vec{k})$, produced in a  process 
of two DM fermions labeled as 1 and 2 scattering to states 
3 and 4 (see Fig.~\ref{fig:diagram} for the Feynmann 
diagram of this process). Each state is defined by the 
4-momentum $p_i^\mu=(E_i,\vec{p}_i)$, where $i$ runs from 1 
to 4. The density number of emittedphotons per unit time 
is given by:
\begin{align}
\frac{d^2N}{dV dt}= d\Pi\gamma&\int d\Pi\vert M\vert^2 f_1 f_2(1-f_3)(1-f_4)(2\pi)^4 \nonumber\\
&\times\delta^{(4)}(p_1+p_2-p_3-p_4-k)\ ,\label{master}
\end{align}
where
\begin{equation}
d\Pi=\prod_{i=1}^4\frac{d^3p_i}{(2\pi)^32E_i}, \quad d\Pi\gamma=\frac{d^3k}{(2\pi)^3 2\omega}, 
\quad f_i=\frac{1}{e^{\frac{E_i-\mu}{T}}+1}\ ,
\label{phasespace}
\end{equation}
with $f_i$ being the Fermi function, and $\mu$ and $T$ the 
DM chemical potential and the star's temperature, 
respectively. For sake of simplicity, we introduce the 
variable $X_i=\frac{E_i-\mu}{T}$, and recast the phase 
space volume element as:
\begin{equation}
d^3p_i=p_i^2dp_i d\Omega_i\simeq	p_\tn{F} p_idp_id\Omega_i,\,\label{d3p}
\end{equation}
where $d\Omega_i$ is the solid angle specified by 
$\vec{p}_i$, and $p_\tn{F}$ is the dark particle Fermi 
momentum, such that $E_\tn{F}=p_\tn{F}^2/(2m_\tn{X})$. 

\begin{figure}[h]
\centering
\includegraphics[width=9cm]{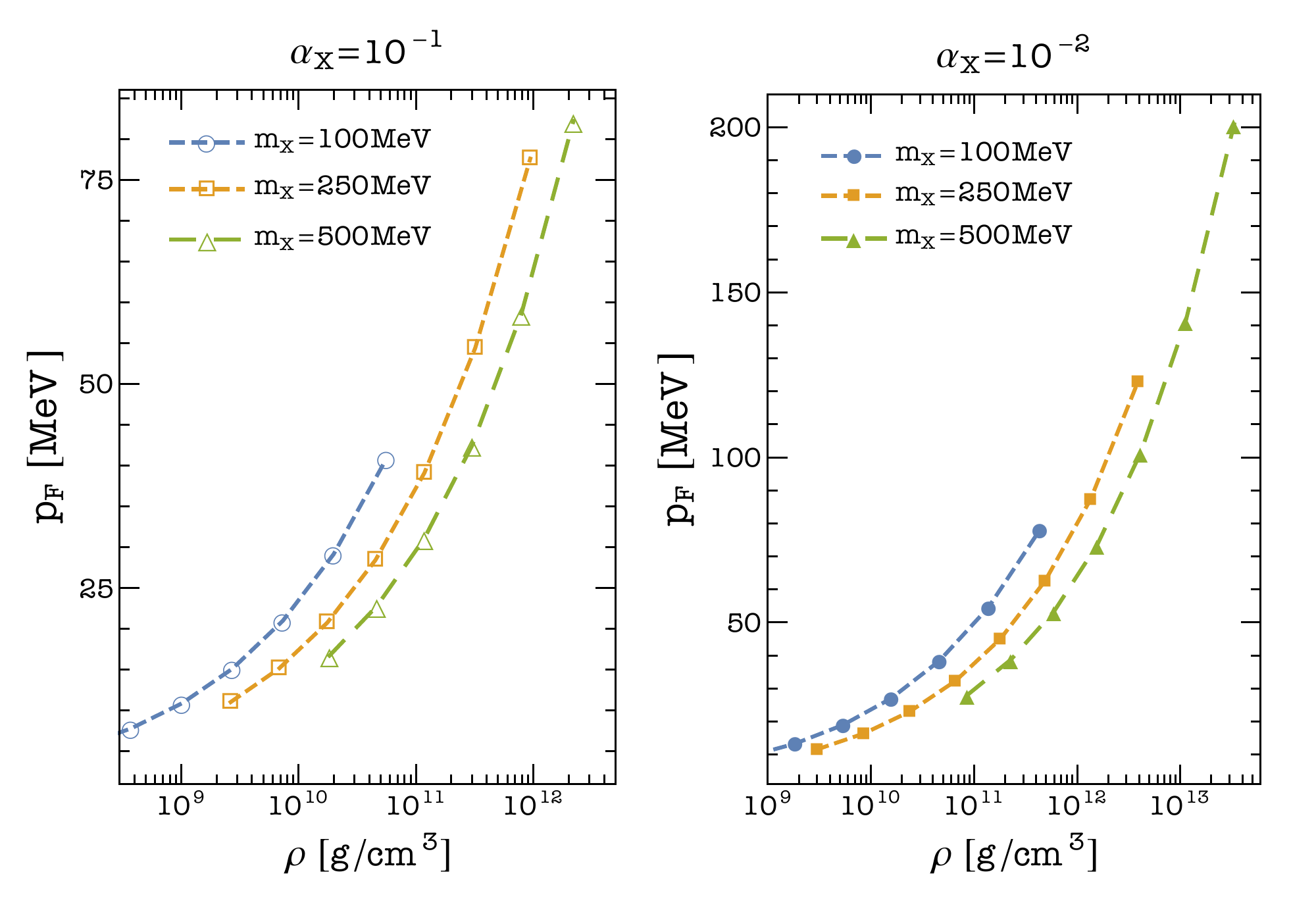}
\caption{The DM Fermi momentum as a function of the 
central energy density for the three different stellar 
configurations shown in Fig.~\ref{fig:MRprofiles}, i.e., 
$m_\tn{X}=(100,250,500)$MeV, and $m_{A'}=10$MeV. The 
right end of the lines corresponds to the central energy 
density of the stellar configuration with the maximum 
mass for fixed DM and dark photon masses as well as 
$\alpha_X$.} 
\label{fig:fermimomentum}
\end{figure}

The latter can be computed from the star's energy density 
profile $\rho$, obtained by solving the relativistic 
stellar structure equations, namely
\begin{equation}
\rho=\frac{g}{(2\pi)^3}\int_0^{\infty} E\frac{d^3p}{e^{\frac{E-\mu}{T}}+1}\simeq
\frac{1}{\pi^2}\int_0^{p_\tn{F}} E p^2dp\ ,\label{fermik}
\end{equation}q
where $E=\sqrt{p^2 +m_\tn{X}^2 }$, and $g=2$ reflects 
the two spin states of the $X$. We assume a system of 
fermions at low temperaturesuch that the system is 
degenerate i.e., the chemical potential is much larger 
than the temperature, $\mu\gg T$. Equation~\eqref{fermik} 
allows to compute the Fermi momentum as a function of the 
radius inside the star, i.e. $p_\tn{F}(r)=p_\tn{F}[\rho(r)]$. 
As an example, in Fig.~\ref{fig:fermimomentum} we show 
the values of $p_\tn{F}$ versus the central energy density 
corresponding to all the stellar configurations displayed 
in Fig.~\ref{fig:MRprofiles}.

Replacing the expressions  \eqref{phasespace}-\eqref{d3p} 
into the master equation \eqref{master} then yields:
\begin{widetext}
\begin{align}
\frac{d^2N}{dV dt}=&\frac{d\Pi\gamma}{16(2\pi)^{8}}p_\tn{F}^4T^4\prod_{i=1}^4\int  d\Omega_i \vert M\vert^2\delta^{(3)}(\vec{p}_1+\vec{p}_2-\vec{p}_3-\vec{p}_4-\vec{k})\times\nonumber\\
&\times\frac{1}{T}\int dX_i \frac{1}{e^{X_1}+1}\frac{1}{e^{X_2}+1}\frac{1}{e^{-X_3}+1}\frac{1}{e^{-X_4}+1}\delta(X_1+X_2-X_3-X_4-\omega/T)\ ,
\end{align}
\end{widetext}
where we have made the assumption that particles are 
nonrelativistic i.e., $E_i\simeq m_\tn{X}$\footnote{The 
assumption is a posteriori verified. As it can be inferred 
from Fig.~\ref{fig:fermimomentum}, the Fermi energy is 
always much smaller than $m_X$.}. Momentum conservation 
requires that $\vec{q}=\vec{p}_1-\vec{p}_3$, and therefore
$q^2=2p_\tn{F}^2(1-\cos\theta_{13})$ ($\theta_{13}$ being 
the angle between $\vec{p}_1$ and $\vec{p}_3$). The 
scattering amplitude can be factorized in terms of the 
elastic scattering amplitude~\cite{Kouvaris:2016afs} as:
\begin{align}
\vert M\vert^2=c&\vert M_{el}\vert^2\frac{1}{m_\tn{X}^2\omega^2}\left(\vec{q}^{\ 2}-\frac{(\vec{q}
\cdot\vec{k})^2}{\omega^2}\right)=\nonumber\\
=c &\vert M_{el}\vert^2\frac{2p_\tn{F}^2}{m_\tn{X}^2\omega^2}(1-\cos\theta_{13})(1-\cos^2\theta_{qk})\ ,
\end{align}
where
 \begin{equation}
c=2(g\epsilon)^2=8\pi\alpha_\tn{X} \epsilon^2\ ,\label{ceq}
\end{equation}
$\theta_{qk}$ the angle between $\vec{q}$ and $\vec{k}$, 
and $g$  the coupling between DM and dark photons 
($\alpha_\tn{X}=g^2/(4 \pi)$). Here $\epsilon$ specifies 
the effective mixing between dark the ordinary photons. 
As mentioned earlier,  the kinetic mixing term of 
eq.~(\ref{eq:lagrangian}) can be formally removed by 
shifting the photon field and therefore in such a case 
no coupling between DM and photons exists. However, in 
the case where there is  a nonzero photon mass inside the dark 
star, the aforementioned shift will remove the kinetic 
mixing term, while introducing a term $\kappa m_D^2A_{\mu}A'^{\mu}$, 
where $m_D$ is the photon mass inside the dark star. Upon 
diagonalization of the mass matrix, there is an induced 
DM-photon coupling with strength $-g\kappa m_D^2/(m_D^2-m_{A'}^2)$. 
Clearly if the photon remains massless inside the dark star, 
no coupling between DM and photons exists. There are (at 
least) three potential scenarios and reasons for a photon 
to have a nonzero mass inside the dark star: i) DM is in 
a degenerate state and the vacuum polarization diagram of 
the photon gets a correction via a loop of DM. This is the 
usual Debye mass contribution which in the case of 
$m_D>m_{A'}$ (and within the hard-dense-loop approximation) 
takes the simple form $m_D=g\kappa \mu/\pi$, where $\mu$ 
is the chemical potential of DM. In this case $\epsilon =\kappa$. 
ii) the existence of even small quantities of protons and 
electrons inside the dark star will also induce a photon 
Debye mass of the order of $e \mu_e/\pi$ where $\mu_e$ 
is the chemical potential of electrons \footnote{Protons 
will also be present to maintain overall electric neutrality.}. 
Despite the fact that the amount of electrons inside the 
star might be negligible to DM, this contribution to the 
Debye mass comes from the direct coupling of photons to 
electrons (or protons) and is not suppressed by $\kappa$. 
iii) Photons might acquire a medium induced mass via a 
Higgs mechanism. One can envision a Higgs-like scalar field 
$\phi$ coupled to photons and DM via an interaction e.g. 
$\phi\bar{X}X$. The nonzero DM density which translates to
a nonzero $\langle \bar{X}X\rangle$ sources in turn a nonzero 
expectation value for $\phi$, effectively providing a mass 
for the photon in the DM infested medium (via a Higgs-like 
mechanism), while photons remain massless outside of the star. 
We leave for future work the precise study of all three 
scenarios. In this paper we are going to assume that the 
photon acquire a mass via any of the above mechanisms and we 
focus on the estimate of the star's luminosity. We should 
mention at this point that as long as the photon  mass is 
smaller than the temperature of the dark star, the kinetic 
mixing will always induce a Bremsstrahlung process to real 
photons.

The factor of 2 in eq.~(\ref{ceq})comes from the fact that 
the photon can be emitted by any of the two DM particles that 
scatter. The amplitude of the elastic process is given by 
\begin{widetext}
\begin{equation}
\vert M_{el}\vert^2=2g^4 \left [\frac{s^2+t^2}{(u-m_{A'}^2)^2}+\frac{s^2+t^2}{(t-m_{A'}^2)^2}+\frac{2s^2}{(t-m_{A'}^2)(u-m_{A'}^2)}\right ],
\label{amp}
\end{equation}
\end{widetext}
where $s=(p_1+p_2)^2$, $t=(p_1-p_3)^2$ and $u=(p_1-p_4)^2$ are the usual Mandelstam variables. To 
proceed and simplify things we will consider two distinct and orthogonal cases. In the first case we assume 
that $m_{A'}>>p_F$, while in the second case we will consider the opposite limit $m_{A'}<<p_F$. In the first 
case the amplitude becomes simple based on the fact that $m_{A'}^2$ will always dominate over $t$ and $u$ 
in the denominators of eq. (\ref{amp}), and since $s\simeq 4m_X^2$, $s>>t$ and $s>>u$, the elastic scattering 
matrix now reads reads:  
\begin{equation}
\vert M_{el}\vert\simeq\frac{64g^4 m_\tn{X}^4}{m_{A'}^4}\ ,\label{Meq}
\end{equation}
where the numerical prefactor comes from 
considering all possible diagrams \cite{berestetskii2012quantum}. A symmetry factor of $1/2$ has been taken 
into account based on the fact that the scattering particles are identical. The previous equation is strictly valid 
when $m_{A'} >> p_\tn{F}$. To obtain this result we have also assumed that the DM particles are non-relativistic 
within the star i.e., $m_X>>p_F$. The rate of photon production is now given by
\begin{align}
\frac{d^2N}{dV dt}
=&\frac{cT^3}{16(2\pi)^{11}}\vert M_{el}\vert^2\frac{p_\tn{F}^6}{m_\tn{X}^2\omega^3}
 I_1 \times I_2\ d^3\omega\ ,
 \label{eq_N}
\end{align}
where we have split the two integrals:
\begin{subequations}
\begin{align}
I_1&=\prod_{i=1}^4\int d\Omega_i \delta^{(3)}(\vec{p}_1+\vec{p}_2-\vec{p}_3-\vec{p}_4)\nonumber\\
&\phantom{aaaaaaaaaaaaaaa}\times (1-\cos\theta_{13})(1-\cos^2\theta_{qk})\ , \\
I_2&=\int  \frac{dX_1}{e^{X_1}+1}\frac{dX_2}{e^{X_2}+1}\frac{dX_3}{e^{-X_3}+1}\frac{1}{e^{-X_1-X_2+X_3+\frac{\omega}{T}}+1}\ .
\end{align}
\label{I12}
\end{subequations}
The integration of $X_i$ is performed within the interval 
$\left[-\frac{\mu}{T}, \infty\right]$; however, since $\mu\gg T $, we can assume 
$\frac{\mu}{T}\rightarrow \infty$, so the integrals can be taken from $\left[-\infty, \infty\right]$. In $I_1$ we omit 
$\vec{k}$ from the delta function since the magnitude of the latter should be of the order of the temperature which 
is much smaller than $\vec{p_i}$ which are practically equal  in magnitude to $p_F$.  In $I_{2}$ we have performed 
the $X_4$ integration using the energy delta function. The integral $I_1$ can be computed analytically. Since all 
$\vec{p}_i$ are on the Fermi surface, the only non-trivial phase space comes from {\it back-to-back} scattering, i.e. 
$\vec{p}_{1}\simeq-\vec{p}_{2}$ and $\vec{p}_{3}\simeq-\vec{p}_{4}$. In this case 
\begin{equation}
I_1=\frac{1}{p_F^3}\int d\Omega_1 d\Omega_3(1-\cos\theta_{13})(1-\cos^2\theta_{qk})=\frac{32\pi^2}{3p^3_\tn{F}}\ ,
\label{I1_value}
\end{equation}
and therefore
\begin{equation}
\frac{d^2N}{dV dt}=\frac{cT^3}{(2\pi)^{9}}\vert M_{el}\vert^2\frac{p_\tn{F}^3}{6m_\tn{X}^2\omega^3}I_2\ d^3\omega\ ,\label{dgamma}
\end{equation}
where $I_2=I_2(\omega)$. 

Combining eqns.~\eqref{ceq}, \eqref{Meq}, and \eqref{dgamma}, we have 
\begin{equation}
\frac{d^2N}{dV dt}=\frac{8}{3}\frac{T^3}{\pi^6} \frac{m_\tn{X}^2}{m_{A'}^4}\frac{p_\tn{F}^3}{\omega^3}\alpha_\tn{X}^3 \epsilon^2 I_2\ d^3\omega\ .\label{dndtdV}
\end{equation}
Given that the photon emission is isotropic and
$d^3\omega=4 \pi \omega^2d\omega$, we can
rewrite  eq.~\eqref{dndtdV} as: 
\begin{equation}
\frac{d^2N}{dV dt}=\frac{32}{3\pi^5}T^3\frac{m_\tn{X}^2}{m_{A'}^4}\frac{p_\tn{F}^3}{\omega}\alpha_\tn{X}^3 \epsilon^2 I_2d\omega\ .\label{dndtdV2}
\end{equation}

In the detector reference frame, time interval and energy 
of emitted photons are redshifted by the source's gravitational 
field, i.e. $\omega_\infty=\sqrt{g_{00}}\omega$ and 
$dt_\infty=dt/\sqrt{g_{00}}$, where the subscript $_\infty$ 
identify redshifted quantities, and $g_{00}$ is the time-time 
component of the metric tensor \emph{inside} the star. Note 
that the latter depends on the radial coordinate $r$, and it 
matches the analytic Schwarzschild expression at the stellar 
surface, where $g_{00}=1-2M/R$. Taking into account the 
redshift and using eq.~\eqref{dndtdV2}, we can obtain the 
flux of photons arriving on an Earth based detector:
\begin{equation}
\frac{d^3N}{dV  dS dt_\infty}=\frac{8}{3}\frac{\sqrt{g_{00}}}{\pi^6}\frac{T^3}{d^2}\frac{m_\tn{X}^2}{m_{A'}^4}
\frac{p_\tn{F}^3}{\omega_\infty}\alpha_\tn{X}^3 \epsilon^2 I_2  d\omega_\infty \ ,
\label{flux11}
\end{equation}
where $dS$ is the detector differential area. To get the flux 
we have divided by a factor of $4 \pi d^2$ ($d$ being the 
distance between Earth and dark star). The reader should keep 
in mind that $g_{00}$ is a function of $r$ the distance from 
the center of the star where the photon was emitted. In general, 
some of the produced photons inside the dark star might not 
make it out, due to the possibility of converting back to dark 
photons via dark Compton scattering. This happens when the mean 
free path of the photon becomes smaller than the
distance it must cross within the star in order to get out. We  take this absorption effect  into account 
 by suppressing the photon flux by  an exponential factor of the optical depth of the stellar 
medium, $\sim e^{-\tau(r)}$, with 
\begin{equation}
\tau(r)=\int_r^R \sqrt{g_{rr}}n_\tn{X}(y) \sigma_\tn{c}dy\ ,\label{optical}
\end{equation}
where $n_\tn{X}$ is the number density of DM particles inside the star, $\sigma_\tn{c}$ is the photon-DM 
scattering cross section which is given by the dark Compton one multiplied by a factor of $\epsilon^2$,  i.e.,
$\sigma_\tn{c}=g^4\epsilon^2/(6\pi m_\tn{X}^2)$, and $g_{rr}$ is the $rr$ component of the metric tensor. By including this factor, we can now integrate over the volume of the whole dark star
\begin{equation}
\frac{d^3N}{dSdt_\infty d\omega_\infty}=\frac{32}{3\pi^5}\frac{T^3}{d^2}\frac{m_\tn{X}^2}{m_{A'}^4}\frac{\alpha_\tn{X}^3 \epsilon^2}{\omega_\infty}\int_{0}^{R}
\frac{\sqrt{g_{00}g_{rr}}}{e^{\tau (r)}}p_\tn{F}^3 I_2\ r^2 dr.\label{numberphot}
\end{equation}

Finally, the total energy flux arriving at the detector ${\cal F}=\frac{d^2E}{dSdt_\infty}$ is:
\begin{align}
{\cal F}
=\frac{32}{3\pi^5}\frac{T^4}{d^2}\frac{m_\tn{X}^2}{m_{A'}^4}\alpha_\tn{X}^3 \epsilon^2\int_{0}^{\infty} I_2dz\int_{0}^{R}\sqrt{g_{00}g_{rr}}
\frac{p_\tn{F}^3r^2}{e^{\tau}}dr,\label{gammaflux}
\end{align}
where we have introduced the dimensionless variable\footnote{Note that $I_2=I_2\left(z/\sqrt{g_{00}}\right)$.} 
$z=\omega_\infty/T$. Recall that $p_F$ is a function of $r$. 

Now we would like to consider the opposite limit $m_{A'}<<p_F$. This is slightly more complicated because the denominator of each of the three terms in eq.~(\ref{amp}) is not simply $\sim m_{A'}^4$ and $M_{el}$ becomes angle depended. In this case the factorization of the angle integrals via the $I_1$ and $I_2$ of Eqs.~(\ref{I12}) is not valid anymore. Recall that at the previously examined  limit  $m_{A'}>>p_F$, $M_{el}$ was independent of angles. At the limit where $m_{A'}<<p_F$ one can easily show that 
the second and third term of eq.~(\ref{amp}) give contributions of the order of $\sim g^4 m_X^4/p_F^4$, while the first one of $\sim m_X^4/(p_F^2 m_{A'}^2)$ which dominates since in this limit $m_{A'}<<p_F$. Therefore we will keep only the contribution from this term in the following derivation.
Within this approximation the amplitude becomes
\begin{equation}
\vert M_{el}\vert^2=\left [16g^4 \frac{m_X^4}{m_{A'}^4} \right ] \frac{1}{(1+\gamma +\gamma \cos \theta_{13})^2},
\label{new_amp}
\end{equation}
where $\gamma= 2p_F^2/m_{A'}^2$. In the above we have taken into account the symmetry factor of $1/2$ due to identical scattering particles. The term within the bracket is the result for this term in the opposite limit i.e., $m_{A'}>>p_F$. Here we are interested in light dark photons i.e., $\gamma>>1$. As mentioned earlier, the amplitude becomes dependent on the angle $\theta_{13}$ and is not constant anymore. This means that the photon production rate will still be given by eq.~(\ref{eq_N}), with $\vert M_{el}\vert^2$ being the angle independent piece of the amplitude i.e., the bracket term of eq.~(\ref{new_amp}) and the angle integral $I_1$ substituted by an $I_1'$ which includes the angle dependent part of eq.~(\ref{new_amp}) and reads
\begin{equation}
 I_1'=\frac{1}{p_F^3}\int d\Omega_1 d\Omega_3\frac{(1-\cos\theta_{13})(1-\cos^2\theta_{qk})}{(1+\gamma +\gamma \cos \theta_{13})^2}.
 \end{equation}
At the limit where $\gamma<<1$, one recovers $I_1'\simeq I_1$ (see eq.~(\ref{I1_value})). The angle $\theta _{qk}$ does not depend on the $\theta_{13}$ and we can simplify things by taking into account that $\overline{\cos^2 \theta_{qk}}=1/3$. Using that, rotating the coordinate system so particle 1 is along the $z-$axis and  performing first the trivial integrations over $\Omega_3$ and $\phi_1$ and then the integration over $\theta_1$, gives to leading order in $1/\gamma$
\begin{equation}
I_1'=\frac{16 \pi^2 m_{A'}^2}{3 p_F^5}.
\label{I1'new}
\end{equation}
As mentioned above, the new version of eq.~(\ref{eq_N}), is with $\vert M_{el}\vert^2$ being the angle independent the bracket term of eq.~(\ref{new_amp}) and the angle integral $I_1$ substituted by $I_1'$. The photon production rate now reads
\begin{equation}
\frac{d^2N}{dV dt}=\frac{T^3}{3\pi^6}\alpha_X^3\epsilon^2\frac{m_X^2}{m_{A'}^2}\frac{p_F}{\omega^3}I_2 d^3\omega,
\label{newdn}
\end{equation}
where we have used eq.~(\ref{ceq}). eq.~(\ref{newdn}) is the new version of eq.~(\ref{dndtdV}) at the limit $m_{A'}<<p_F$. Performing the trivial angular integral of $d^3\omega$ and dividing by $4 \pi d^2$ as in eq.~(\ref{flux11}) we get 
\begin{equation}
\frac{d^3N}{dV  dS dt_\infty}=\frac{T^3}{3 \pi^6 d^2}\sqrt{g_{00}}\alpha_X^3\epsilon^2 \frac{m_X^2}{m_{A'}^2}\frac{p_F}{\omega_{\infty}}I_2d\omega_{\infty}.
\end{equation}
To complete the set of equations,  the $m_{A'}<<p_F$ limit of  Eqs.~(\ref{numberphot}) and (\ref{gammaflux}) are given by
\begin{equation}
\frac{d^3N}{dSdt_\infty d\omega_\infty}=\frac{4}{3\pi^5}\frac{T^3}{d^2}\frac{m_\tn{X}^2}{m_{A'}^2}\frac{\alpha_\tn{X}^3 \epsilon^2}{\omega_\infty}\int_{0}^{R}
\frac{\sqrt{g_{00}g_{rr}}}{e^{\tau (r)}}p_\tn{F} I_2\ r^2 dr
\end{equation}
and 
\begin{align}
{\cal F}
=\frac{4}{3\pi^5}\frac{T^4}{d^2}\frac{m_\tn{X}^2}{m_{A'}^2}\alpha_\tn{X}^3 \epsilon^2\int_{0}^{\infty} I_2dz\int_{0}^{R}\sqrt{g_{00}g_{rr}}
\frac{p_\tn{F}r^2}{e^{\tau}}dr.\label{gammaflux2}
\end{align}



\section{Numerical results}\label{Sec:results}
\begin{figure}[thbp]
\centering
\includegraphics[width=7.cm]{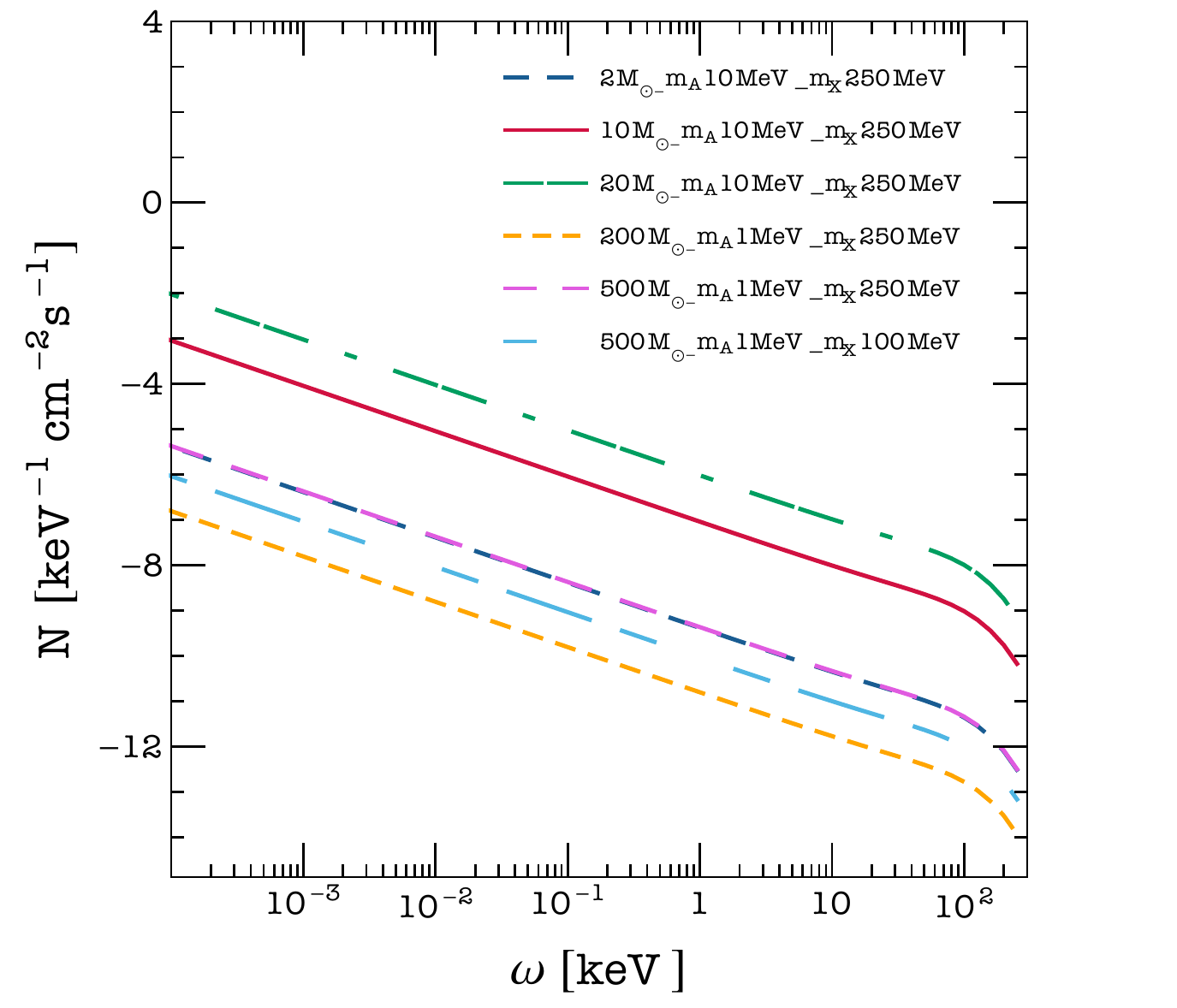}
\caption{Flux of received photons  as a function of 
the energy for different dark stars at a distance 
$d=1$kpc, with temperature of $T=5 \times 10^8$K. 
Coloured curves refer to different values of the dark 
particle, dark photon and stellar masses. For 
$m_A'=10$MeV ($m_A'=1$MeV) we fix $\kappa=10^{-4}$ 
($\kappa=10^{-8}$).}  
\label{fig:numberphot}
\end{figure}

Based on the discussion above we are now in position 
to estimate the photon flux and luminosity that can 
be produced by asymmetric dark stars. At first sight,
what controls the photon production is the coefficient 
$\epsilon$. Clearly too small values of $\epsilon$ 
will lead to undetectable photon signal. As we 
mentioned earlier an induced photon mass will lead 
to an $\epsilon= \kappa m_D^2/(m_D^2-m_{A'}^2)$. The 
kinetic mixing $\kappa$ is constrained 
experimentally~\cite{Jaeckel:2013ija,Bauer:2018onh}. 
For heavy dark photon masses e.g. of the order of 
$m_{A'}=1$GeV, $\kappa<10^{-5}$. Despite the fact that 
smaller dark photon masses are more constrained, they 
can provide a much larger luminosity in dark stars. 
The reason is twofold.  Firstly the mass of the dark 
photon appears in the denominator of the flux equation 
(see e.g. eq.~(\ref{gammaflux2})) and therefore lighter 
dark photons produce larger luminosities. The second 
reason is also that light dark photons (for example 
lighter than the medium acquired photon mass $m_D$) 
saturate $\epsilon$ to $\kappa$ providing significant 
amount of mixing\footnote{In fine tuned situations 
where $m_D\simeq m_{A'}$ $\epsilon$ can be become in 
principle much larger than $\kappa$. We will not 
examine such fine tuned cases here.}. The value of 
$\epsilon$ or $\kappa$ can also affect the photon production 
in two different ways: generally large values of 
$\epsilon$ correspond to larger photon production since 
the latter scales as $\sim \epsilon^2$. However in 
principle, too strong $\epsilon$ can allow re-conversion 
of produced photons at the center of the star back to 
dark photons due to large cross section of a dark 
Compton process. If photons produced at the center are 
reabsorbed by the star, then the emitted photons will 
come from layers close to the surface leading to an 
overall reduction.  This represents a major difference 
with respect to the emission from standard NS where the 
bulk of the star does not contribute due to the very 
short photon mean free path inside the core. As a result, 
the spectrum of photons produced via Bremsstrahlung 
inside dark stars is qualitatively different from that 
of NS. In the latter case, one should expect the usual 
black-body spectrum (sometimes deformed by the star's 
atmosphere) peaking roughly at a frequency close to the 
surface temperature of the star. On the contrary, the 
potential spectrum from dark stars is not solely produced 
from photons produced on the surface of the star, but it 
rather comes from the whole bulk redshifted appropriately 
depending on the depth produced. Therefore although the 
photon spectrum in dark stars is $\epsilon^2$ suppressed, 
this is to some extent counterbalanced by the fact that 
the whole volume of the star participates in the photon 
production. In addition, the fact that the produced photons 
are not in thermal equilibrium with the medium leads to 
a different  shape in the spectrum compared to photon 
production from the surface of a NS where the photons are 
always in thermal equilibrium with the nuclear matter 
and therefore the spectrum is that of black-body. The 
photon spectrum from dark stars in shown 
Fig.~\ref{fig:numberphot}, where we plot the flux of 
received photons per energy, namely 
$d^3N/(dS dt_\infty d\omega_\infty)$ in 
eq.~\eqref{numberphot}. For energies smaller than 
$\sim T$, the spectrum is determined by the characteristic 
$\omega_{\infty}^{-1}$ of the Bremsstrahlung rate, 
while the rate drops faster above $\sim T$ due to a 
thermal exponential suppression. Overall, we have 
tried different parameter values for heavy dark photons 
within the allowed range of $\kappa$ and despite the 
fact that the star is not opaque to the produced photons 
i.e., the mean free path of the photons is much larger 
than the size of the star and therefore all photons make 
it out from the star, the produced photon spectrum 
is too small to be detected within current observational 
capabilities. As shown in Fig.~\ref{fig:numberphot} the 
situation changes once we consider light dark photons, 
because as we mentioned not only they lead to larger 
mixing with the ordinary photons, but they also give 
an overall higher amplitude. 

\begin{figure*}[thbp]
\centering
\includegraphics[width=7.5cm]{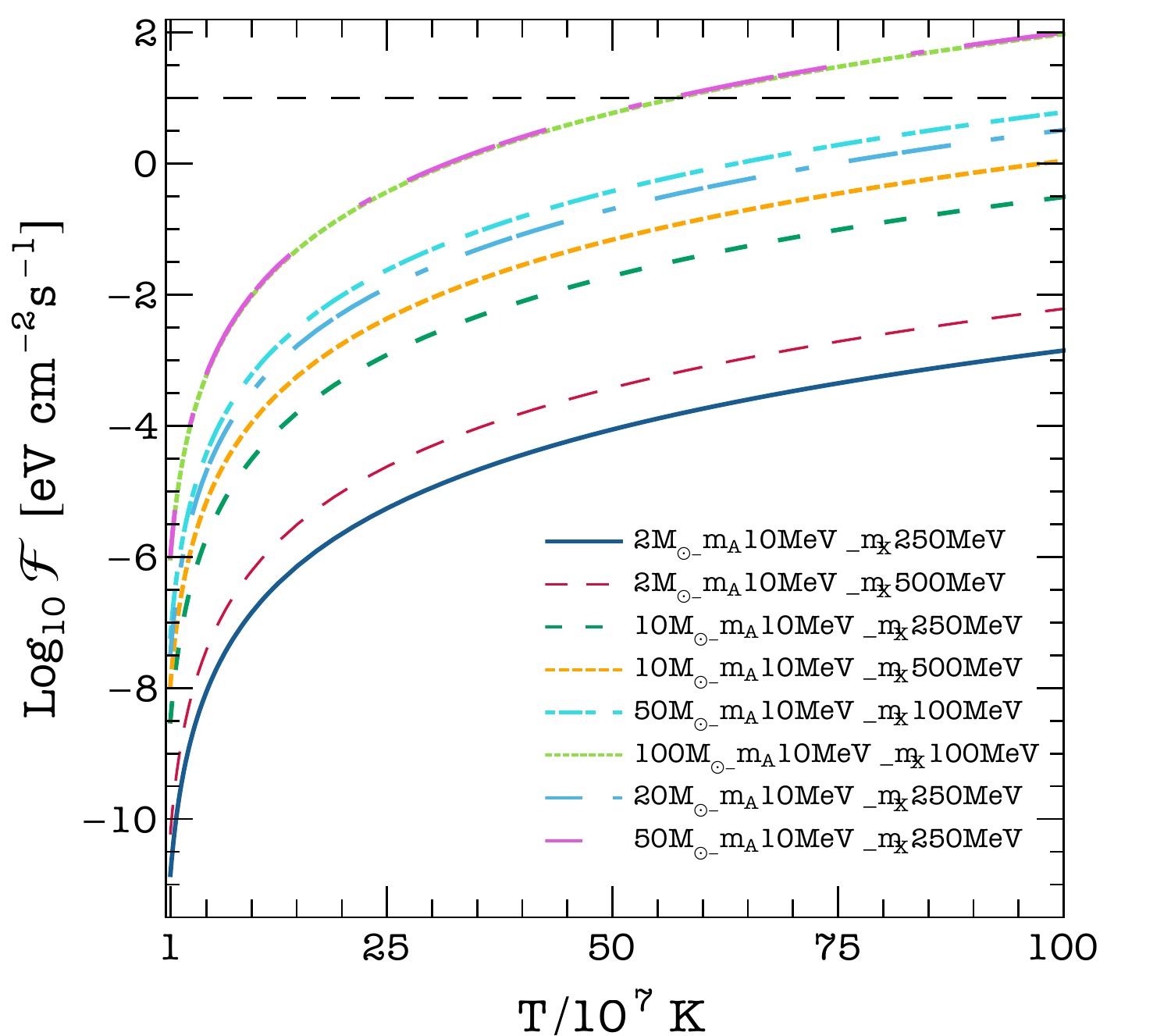}
\includegraphics[width=7.5cm]{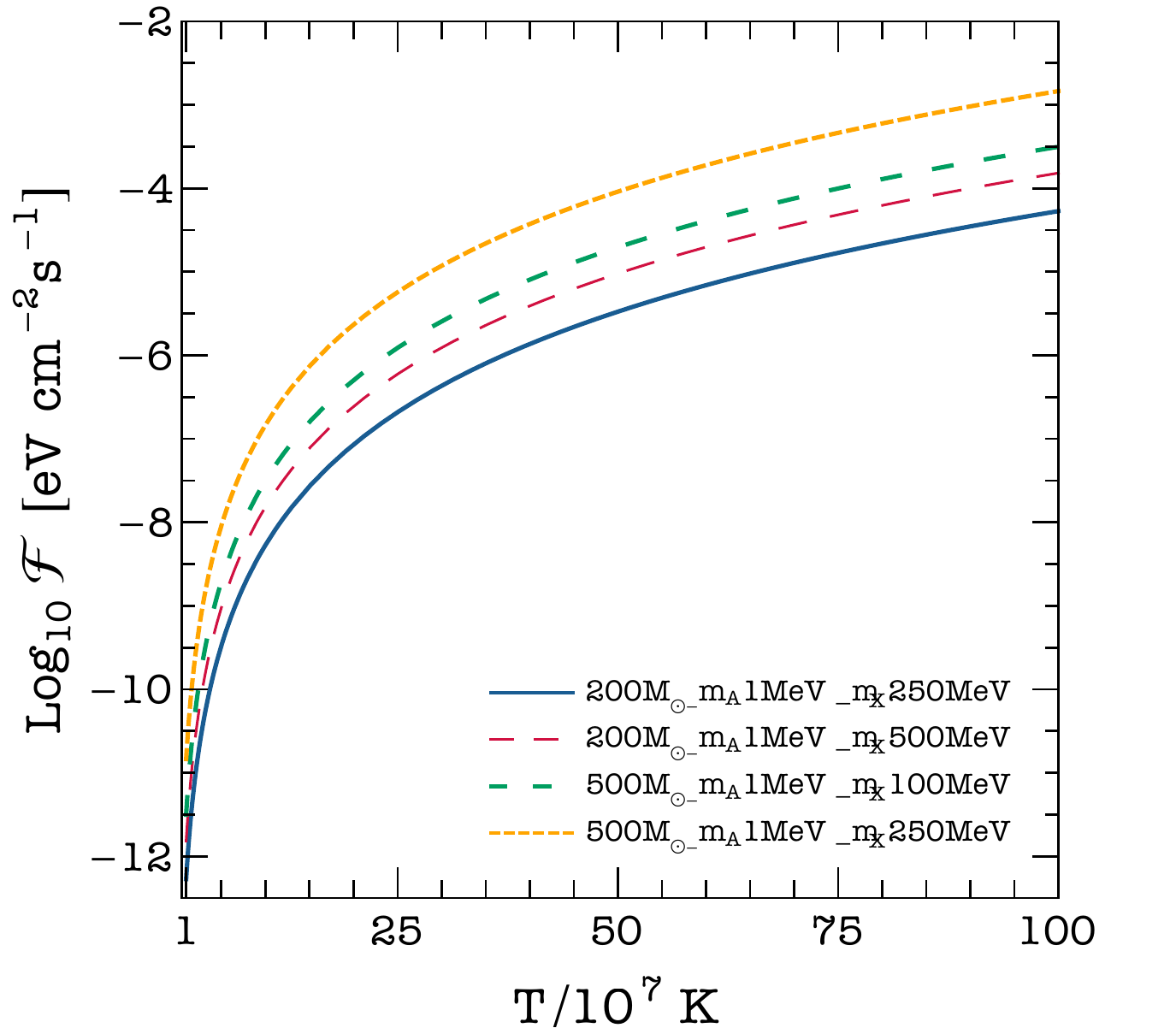}
\caption{Bremsstrahlung photon energy flux produced by 
dark stars located at a distance $d=1$kpc away from the 
detector with $m_\tn{X}=(100,250,500)$MeV and $m_{A'}=10$MeV 
(left panel), $m_{A'}=1$MeV (right panel), as a function 
of the star's temperature. We consider different values of 
the stellar mass, fixing $\kappa=10^{-4}$ for $m_{A'}=10$MeV, 
$\kappa=10^{-8}$ $m_{A'}=1$MeV, and $\alpha_\tn{X}=10^{-1}$. 
The dashed horizontal line corresponds to 
${\cal F}=10$eV cm$^{-2}$s$^{-1}$.} 
\label{fig:flux}
\end{figure*}

Figures~\ref{fig:flux} show the photon energy flux 
\eqref{gammaflux} produced by a large variety of 
dark stars with various masses and EoS. We consider 
sources at a prototype distance of $d=1$kpc. However, 
since ${\cal F}$ is proportional to $1/d^2$, these 
results can be immediately  rescaled to any location. 
In Fig.~\ref{fig:flux} in particular we draw ${\cal F}$ 
as a function of the normalised temperature $T_7=T/10^7$K 
for dark stars with $m_A'=10$MeV and $m_A'=1$MeV. 
We use two prescriptions for $\kappa$, choosing 
$\kappa=10^{-4}$ and $\kappa=10^{-8}$ for the heavy 
and the light dark photon, respectively. The latter 
plays a key role in determining the amplitude of the flux, 
which is overall favoured by small values of $m_A'$. 
Note that ${\cal F}$ in eq.~\eqref{gammaflux2} is 
proportional to $T^4$, so these results can be 
extrapolated to any temperature by the appropriate 
rescaling. For $T\lesssim 5\times 10^8$K all the stellar 
models considered lead to values of ${\cal F}$ smaller 
than $1$eV cm$^{-2}$s$^{-1}$. The Bremsstrahlung fluxes 
grow with the temperature, and can be as high as 
${\cal F}\sim 10^{2}$eV cm$^{-2}$s$^{-1}$ for massive 
stellar configurations with $M\gtrsim50M_\odot$ and 
$T=10^9$ K. Although the flux in eq.~\eqref{gammaflux2} 
depends  directly on the mass of the DM particles as 
${\cal F}\sim m_\tn{X}^2$, it also depends indirectly 
on $m_\tn{X}$ because the latter affects the stellar 
density profiles, namely the stiffness of the fermion 
star. Indeed for fixed particle parameters (i.e., DM and 
dark photon mass, $\kappa$ and  coupling $\alpha_X$), 
more massive stars have higher luminosities.

At this point we can make a comparison between the 
luminosity of an asymmetric dark star with that of a NS. 
Assuming a thermally cooling weakly magnetised NS, the 
flux produced at the surface is proportional to the 
stellar surface temperature, ${\cal F}_\tn{NS}=\sigma_\tn{B} T_\tn{sur}^4R^2/d^2$, where $\sigma_\tn{B}$ is the Stefan-Boltzmann constant 
and $T_{\tn{sur}}$ the surface temperature of the NS. 
A typical NS of $1.5 M_\odot$ and $R\sim11$km at 
$d=1$kpc, with a temperature within the range 
$[10^7, 10^9]\text{K}$,  provides a flux of 
${\cal F}_\tn{NS}\in[10^5,10^{13}]$eV cm$^{-2}$s$^{-1}$ 
on Earth. As we already mentioned the luminosity of a 
dark star scales also as $T^4$ (see eq.~\eqref{gammaflux2}). 
Therefore for given DM parameters and mass of the dark 
star, its photon energy flux will always be a specific 
fraction of that of a typical NS with the same temperature 
located at the same distance from the Earth. On the other 
hand, larger values of $\kappa$, or smaller distances 
may change this picture. 
Let's take for example the two pulsars J0437-4715 and 
J0108-1431, which are about 140pc and 130pc away from 
the Earth, and have surfaces temperatures of 
$\sim 10^5$K~\cite{Kargaltsev:2003eb,Mignani:2008jr}. 
All the dark stars shown in the left panel of 
Fig.~\ref{fig:flux} with $M\gtrsim 50M_\odot$ at 
the same distance, would produce 
a flux higher than J0437-4715 and J0108-1431, provided 
that their temperature is larger than $T\sim10^7K$.

A couple of  comments are in order here. NS evacuate 
energy from the bulk via neutrino emission. In fact the 
modified Urca process is the main mechanism of NS cooling 
that dominates over surface photon emission in temperatures 
above $\sim 10^8$K. A process that would be somewhat analogous 
to our dark Bremsstrahlung is the neutrino Bremsstrahlung 
emission inside a NS studied in \cite{Jaikumar:2005gm}. However 
this process scales parametrically differently with temperature 
from our case, since they are two thermal particles produced 
(a pair of neutrino-antineutrino) instead of one (the photon) 
in our case. A second point is that the dark star luminosity 
(as it can be seen in Fig. \ref{fig:flux})  can be significantly 
higher if the value of $\alpha_X$ is larger. 
For a given DM and dark photon mass, the value of $\alpha_X$ 
has an upper bound  due to constraints on DM self-interactions 
from the bullet cluster. However, this constraint does not apply 
if  $X$ does not account for the whole DM abundance but only 
a fraction  of it.


\section{Conclusion}\label{Sec:conclusion}

Asymmetric DM is an attractive alternative to the thermally 
produced WIMP paradigm. Due to an inherited asymmetry between 
particles and antiparticles, DM annihilations are absent once the population of antiparticles is depleted. Therefore in case  such a 
DM candidate possesses an effective mechanism of evacuating energy, it has been recently demonstrated~\cite{Chang:2018bgx} 
that such a DM component could collapse and form compact objects. In fact this is also desired in the view of the supermassive 
black holes which seem in need of seeds other than typical stellar remnants.

Up to now, it was considered that asymmetric dark stars could be detected only via gravitational wave signals. 
In fact in previous work, we had investigated and computed the tidal deformabilities of such stars and the prospects of their discovery 
via gravitiational waves~\cite{Maselli:2017vfi}. 
In this 
paper we demonstrate that in some cases dark stars might also be detected by direct  observation of a photon spectrum produced 
inside such a star via a dark Bremsstrahlung provided that dark photons mix kinetically with ordinary photons.
We calculated the explicit form of the photon flux in terms of the stellar and EoS parameters. We numerically computed 
the emitted flux for a variety of model parameters. Obviously the overall process is strongly affected by the distance between the Earth 
and the dark star and by the kinetic mixing. 
We  found that the numbers of emitted photons is enhanced for heavier/larger dark stars. Depending on the stellar temperature, the Bremsstrahlung flux can be as high as 100 eV cm$^2$s$^{-1}$ for galactic sources. Although this process is in general smaller than the 
energy emitted by standard NS due to black-body radiation, the 
dark photon spectrum features a spectrum morphology, which is completely different from a thermal black-body component, thus 
providing a distinct discovery signature for dark stars. Finally it will be interesting to determine the rate of cooling for such asymmetric 
dark stars which will be dominated by emission (via dark Bremsstrahlung) of dark photons. In case where DM is admixed with baryonic matter and electrons, it will also be interesting to study how these particles affect the overall cooling and luminosity of the dark star. 
 In addition, compact asymmetric dark stars can be composed not only of fermionic DM as in the case studied here but also of bosonic DM (e.g. studied in~\cite{Eby:2015hsq}). In this case the Bremsstrahlung mechanism can work in a similar way as in our current study. However, the emissivity will be different due to the fact that in the bosonic case there is no Fermi surface. In our present study only DM particles sitting within $\sim T$ from the Fermi surface can interact and produce photons. This Pauli blocking is not present in the case of bosonic stars and therefore the amount of produced photons could potentially be larger than the one studied here. Therefore the calculation is completely different. 
 We will address all the above in future work. \\
\noindent{\em Acknowledgements.---}
We would like to thank M. Tytgat for useful discussions. A.M. acknowledge support from the Amaldi Research Center funded by the MIUR 
program ``Dipartimento di Eccellenza'' (CUP: B81I18001170001). CK is partially funded by 
the Danish National Research Foundation, grant number DNRF90, and by the Danish Council 
for Independent Research, grant number DFF 4181-00055. A.M. and K.K 
 would like to acknowledge networking support by the COST Action CA16214.

\bibliography{bibnote}

\end{document}